# DUST GRAIN ORBITAL BEHAVIOR AROUND CERES

R. Nazzario, T. W. Hyde, and L. Barge

*CASPER (*Center for Astrophysics, Space Physics and Engineering Research)
*Baylor University, P.O. Box 97310, Waco, TX 76798-7310, USA*

## ABSTRACT

Many asteroids show indications they have undergone impacts with meteoroid particles having radii between 0.01 m and 1 m. During such impacts, small dust grains will be ejected at the impact site. The possibility of these dust grains (with radii greater than $2.2 \times 10^{-6}$ m) forming a halo around a spherical asteroid (such as Ceres) is investigated using standard numerical integration techniques. The orbital elements, positions, and velocities are determined for particles with varying radii taking into account both the influence of gravity, radiation pressure, and the interplanetary magnetic field (for charged particles). Under the influence of these forces it is found that dust grains (under the appropriate conditions) can be injected into orbits with lifetimes in excess of one year. The lifetime of the orbits is shown to be highly dependent on the location of the ejection point as well as the angle between the surface normal and the ejection path. It is also shown that only particles ejected within 10° relative to the surface tangential survive more than a few hours and that the longest-lived particles originate along a line perpendicular to the Ceres-Sun line.

## INTRODUCTION

Observations of asteroids such as Eros and Mathilde indicate an environment where significant collisions have occurred in the past (Chapman et al., 1999, Davis 1999, Veverka et al., 1999). These collisions would have created debris, much of it in the form of dust particles. Experimental hypervelocity impact data suggests that the smaller debris particles (those with radii less than 1 mm in radius, henceforth called dust) have velocity distributions which might enable them to enter either short- or long-term orbits about the asteroid (Nakamura and Fujiwara, 1991). The dynamics of such ejecta is important in constructing collisional time-scales for dust present in asteroidal environments since the dust can adversely affect spacecraft operations in a variety of ways. The possibility of these dust grains (with radii greater than $2.2 \times 10^{-6}$ m) forming a halo around a spherical asteroid (such as Ceres) is investigated using standard numerical integration techniques. For this study, the asteroid Ceres was chosen as a test case both for modeling simplicity and the increased probability (because of its large mass) that it would retain dust in stable orbits.

## NUMERICAL MODEL

Time-dependent orbits were calculated for each particle primarily utilizing a fifth order Runge-Kutta method (Nazzario and Hyde, 1997, Nazzario and Hyde, 2002, Nazzario, 2002) based on the Butcher's scheme (Chapra and Canale, 1985). The Runge-Kutta method utilized a fixed time step and required six function evaluations for each step. Runge-Kutta methods are well suited to first order differential equations allowing the problem (since positions are dependent on accelerations, i.e. a second order differential equation) to be broken down into two first order differential equations of the general form

$$y_{i+1} = y_i + \phi(x_i, y_i, h)h. \qquad (1)$$

In Eq. (1) h is the step size and the function φ is known as the increment function and has the general form

$$\phi = a_1 k_1 + a_2 k_2 + \ldots + a_n k_n. \quad (2)$$

In Eq. (2) the a's represent constants and the k's follow a recursive relationship given by

$$k_n = f(x_i + p_{n-1}h, y_i + q_{n-1,1}k_1 h + q_{n-1,2}k_2 h + \ldots + q_{n-1,n-1}k_{n-1}h). \quad (3)$$

Since the k's follow a recursive relationship, this situation is ideally suited for solution using a computer. In such iterative methods there is always at least one free parameter which must be chosen. The specific number of free parameters will depend on the order of the Runge-Kutta method but is typically one. Once this free parameter is chosen, a Taylor series expansion can then be used to calculate the a's, p's, and q's. This calculation is accomplished by setting the equations equal to the terms in a Taylor series expansion given by

$$y_{i+1} = y_i + f(x_i, y_i)h + \frac{f'(x_i, y_i)}{2}h^2 + \ldots + \frac{f^{n-1}(x_i, y_i)}{n!}h^n + O(h^{n+1}). \quad (4)$$

For Butcher's scheme this becomes (Chapra and Canale, 1985)

$$y_{i+1} = y_i + \frac{h}{90}[7k_0 + 32k_2 + 12k_3 + 32k_4 + 7k_5]. \quad (5)$$

The $k_i$ values are given by

$$k_0 = f(t_i, y_i, v_i) \quad (6)$$

$$k_1 = f\left(t_i + \frac{1}{4}h, y_i + \frac{1}{4}hk_0, v_i + \frac{1}{4}hM_0\right) \quad (7)$$

$$k_2 = f\left(t_i + \frac{1}{4}h, y_i + \frac{1}{8}hk_0 + \frac{1}{8}hk_1, v_i + \frac{1}{8}hM_0 + \frac{1}{8}hM_1\right) \quad (8)$$

$$k_3 = f\left(t_i + \frac{1}{2}h, y_i - \frac{1}{2}hk_1 + hk_2, v_i - \frac{1}{2}hM_1 + hM_2\right) \quad (9)$$

$$k_4 = f\left(t_i + \frac{3}{4}h, y_i + \frac{3}{16}hk_0 + \frac{9}{16}hk_3, v_i + \frac{3}{16}hM_0 + \frac{9}{16}hM_3\right) \quad (10)$$

$$k_5 = f\left(t_i + h, y_i - \frac{3}{7}hk_0 + \frac{2}{7}hk_1 + \frac{12}{7}hk_2 - \frac{12}{7}hk_3 + \frac{8}{7}hk_4, v_i - \frac{3}{7}hM_0 + \frac{2}{7}hM_1 + \frac{12}{7}hM_2 - \frac{12}{7}hM_3 + \frac{8}{7}hM_4\right) \quad (11)$$

The k's are related to the velocities while the M's correspond to the accelerations at the calculation points. Since the change in velocity is not explicitly dependent on either position or time, the above equations can be reduced to:

$$k_0 = v_i \quad (12)$$

$$k_1 = v_i + \frac{1}{4}hM_0 \quad (13)$$

$$k_2 = v_i + \frac{1}{8}hM_0 + \frac{1}{8}hM_1 \quad (14)$$

$$k_3 = v_i - \frac{1}{2}hM_1 + hM_2 \quad (15)$$

$$k_4 = v_i + \frac{3}{16}hM_0 + \frac{9}{16}hM_3 \quad (16)$$

$$k_5 = v_i - \frac{3}{7}hM_0 + \frac{2}{7}hM_1 + \frac{12}{7}hM_2 - \frac{12}{7}hM_3 + \frac{8}{7}hM_4. \quad (17)$$

Combining terms, Eq. (5) can be reduced to

$$y_{i+1} = y_i + \frac{h}{90}\left[90v_i + 7hM_0 + 24hM_2 + 6hM_3 + 8hM_4\right]. \quad (18)$$

Employing the same procedure to solve for the new velocity yields

$$v_{i+1} = v_i + \frac{h}{90}\left[7M_0 + 32M_2 + 12M_3 + 32M_4 + 7M_5\right]. \quad (19)$$

In the above, the M's are given by required function evaluations, and like the k's, follow a recursive formulation. They are coupled to the position and the velocity in the following manner:

$$M_0 = g(t_i, y_i, v_i) \quad (6)$$

$$M_1 = g\left(t_i + \frac{1}{4}h,\ y_i + \frac{1}{4}hk_0,\ v_i + \frac{1}{4}hM_0\right) \quad (7)$$

$$M_2 = g\left(t_i + \frac{1}{4}h,\ y_i + \frac{1}{8}hk_0 + \frac{1}{8}hk_1,\ v_i + \frac{1}{8}hM_0 + \frac{1}{8}hM_1\right) \quad (8)$$

$$M_3 = g\left(t_i + \frac{1}{2}h,\ y_i - \frac{1}{2}hk_1 + hk_2,\ v_i - \frac{1}{2}hM_1 + hM_2\right) \quad (9)$$

$$M_4 = g\left(t_i + \frac{3}{4}h,\ y_i + \frac{3}{16}hk_0 + \frac{9}{16}hk_3,\ v_i + \frac{3}{16}hM_0 + \frac{9}{16}hM_3\right) \quad (10)$$

$$M_5 = g\left(t_i + h,\ y_i - \frac{3}{7}hk_0 + \frac{2}{7}hk_1 + \frac{12}{7}hk_2 - \frac{12}{7}hk_3 + \frac{8}{7}hk_4,\ v_i - \frac{3}{7}hM_0 + \frac{2}{7}hM_1 + \frac{12}{7}hM_2 - \frac{12}{7}hM_3 + \frac{8}{7}hM_4\right) \quad (11)$$

After calculating the M's for the current position and velocity, the new position and velocity can then be calculated.

A secondary algorithm utilizing a variable time-step was also employed to help determine an appropriate step size for the fifth order method. This algorithm, known as the Runge-Kutta-Fehlberg method (Danby 1997), is a fifth order variable time-step method utilizing two Runge-Kutta formulations (a fourth- and a fifth-order) to estimate the local truncation error. If the local truncation error exceeded a preset tolerance ($\delta$) then a new time-step would be calculated. This works well for simulations of this type as the majority of the force involved is due to the gravitational acceleration of Ceres and the dust particles do not interact with each other. A tolerance of $1 \times 10^{-6}$ was utilized in Eq. (26) (Danby 1997) to calculate the new time step which resulted in changes in energy, semi-major axis, and eccentricity of approximately $2 \times 10^{-8}$ in a one-year simulation using time steps of 30 seconds. Therefore, a time step of 30 seconds was utilized in the main program and then verified by tracking one test particle for a period of 100 years. No changes greater than $2 \times 10^{-6}$ were observed for energy, angular momentum, and inclination (conserved values) during the test run.

$$h_{new} = 0.9h\left(\frac{\delta}{|y_{fifth} - y_{fourth}|}\right)^{1/5} \quad (26)$$

The model was also utilized to obtain numerical solutions for the accelerations acting on individual dust particles incorporating the various forces acting upon them. For this simulation, Ceres' gravity (assuming a spherical contribution only), solar gravity, the solar radiation and Poynting-Robertson effects, and the force created by the interaction of the interplanetary magnetic field with the charged particle were all taken into consideration.

The acceleration on the particle due to the gravitational force is given by

$$\vec{a}_{Grav} = -\frac{GM_C}{r^2}\hat{r} + GM_{Sun}\left[-\frac{\vec{r}}{R^3} + \vec{\rho}\left(\frac{1}{R^3} - \frac{1}{\rho^3}\right)\right]. \qquad (27)$$

In Eq. (27), G is the gravitational constant, $M_C$ is the mass of Ceres, R is the distance from the dust grain to the Sun, $\vec{\rho}$ is the vector from Ceres to the Sun, and $\vec{r}$ is the position vector from the center of Ceres to the dust grain. The first term of Eq. (27) is from Ceres while the second term is that from the Sun taking into account the non-inertial reference frame centered on Ceres (Danby, 1992).

The acceleration due to the radiation pressure is given by (Burns et al., 1979)

$$\vec{a}_{Radiation} = \frac{\beta GM_{Sun}}{R^2}\hat{R} \qquad (28)$$

where $M_{sun}$ is the mass of the Sun and $\vec{R}$ is the vector from the dust grain to the Sun. $\beta$ is defined as

$$\beta = \frac{.6Q}{a\rho_d} \qquad (29)$$

with Q being the radiation pressure efficiency, a the radius of the particle and $\rho_d$ the density of the particle. Since in this work all particles are assumed to have radii greater than $1.0 \times 10^{-6}$ m, the radiation pressure efficiency is taken to be 1.0 and $\rho_d$ is assumed to be the bulk density of Ceres (2800 kg/cm$^3$).

The interplanetary magnetic field will influence charged particles producing a corresponding acceleration of

$$\vec{a}_{Magnetic} = q\vec{v} \times \vec{B}. \qquad (30)$$

In Eq. (30), $\vec{B}$ is the interplanetary magnetic field, $\vec{v}$ is the velocity of the dust grain relative to the magnetic field and q is the charge on the particle. Ejected dust particles should quickly charge to values corresponding to those found in interplanetary space thus simplifying the charging calculation. In this study, the charge on the dust grain was taken to be 3 V upon ejection which is the calculated potential for a dust grain in the interplanetary medium (Kimura and Mann, 1998).

## INITIAL CONDITIONS

A total of six impact sites were selected on Ceres, located at surface points along each of the six axes as shown in Figure 1. Ceres was initially positioned at aphelion (Lang, 1992) revolving around the Sun in the counter-clockwise direction. Each impact was assumed to eject a total of 72,000 particles. Since any single impact will produce a variety of ejecta sizes and speeds, these particles were assumed to be ejected at different speeds and angles relative to the surface normal as shown in Figure 2. 100 particles of differing sizes (ranging from 2.14 $\times 10^{-6}$ m to 214$\times 10^{-6}$ m) were ejected at speeds between 430-600 m/s in increments of 10 m/s. This speed distribution range was chosen to agree with results found by Nakamura (1992). Each set of particles was ejected at angles between 0° to 90° relative to the surface normal (Figure 2) with a 10° increment between ejection angles. The ejection direction was varied along the possible axes ($\pm$x, $\pm$y, and $\pm$z) at each point resulting in a total of 72,000 particles. Thus, a total of 432,000 particles were examined. Upon ejection, the particles were subjected to the forces listed above. As shown in Figure 3, the shadowing effect of Ceres was taken into account for the radiation pressure force whenever the dust particle was in the asteroid's shadow.

## RESULTS

As shown in Table 1, particles launched at angles less than 80° from the asteroid's surface normal were quickly removed from the system either by collision with Ceres or by escaping to the interplanetary media. Only particles ejected within 10° of the surface tangent achieved stable orbits. Of the 432,000 particles launched, 3,504

(0.8%) survived for one Earth year. Of those, particles launched from the poles were most likely to survive with 2916 (83.2% of the surviving particles) of 14,400 launched from the poles entering stable orbits about Ceres.

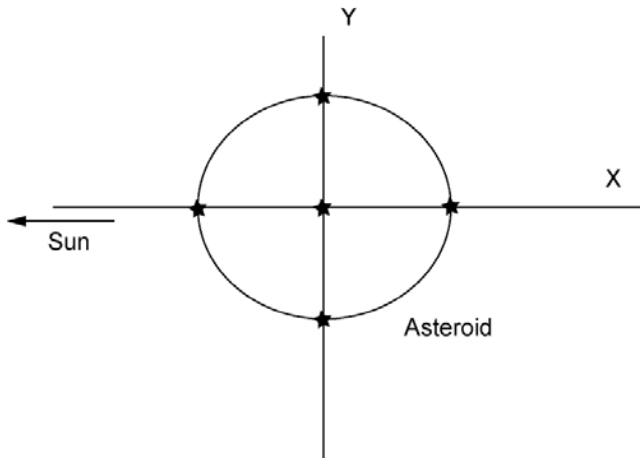

Fig. 1. Ceres' frame of reference. Impact points are denoted by the five stars with the sixth impact point opposite the center star. The z-axis is into the page.

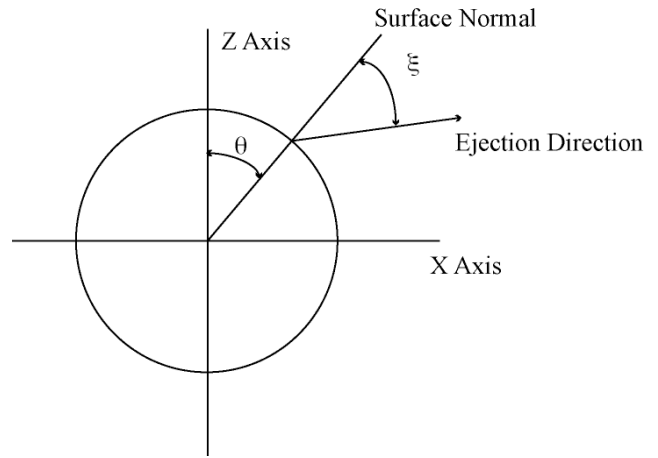

Fig. 2. A two-dimensional drawing showing the ejection angle $\xi$, relative to Ceres' surface normal. The angle $\xi$ represents the opening of a cone in three-dimensions centered on the surface normal.

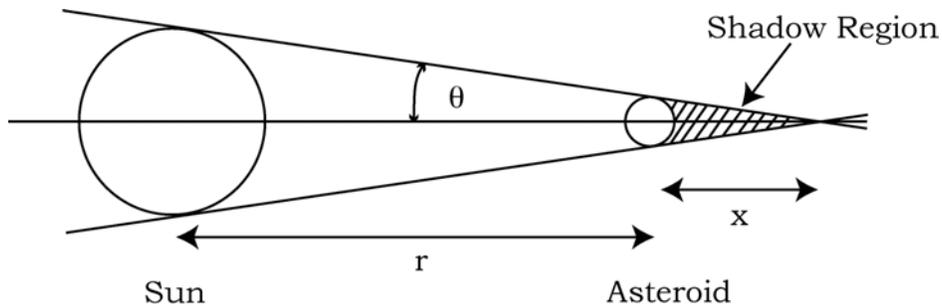

Fig. 3. Asteroid shadowing effect showing the asteroid, the Sun and the shadow region.

Table 1. Particle launch point data. The number of particles surviving, the average radii of the surviving particles and the average final speed are shown after 1 year of simulation time.

| Launch Point | Direction of Launch | Number of Particles Surviving | Average Radii ($\times 10^{-6}$ m) | Average Speed (m/s) |
|---|---|---|---|---|
| X | −y | 97 | 53 | 214.55 |
| Y | −x | 106 | 51 | 246.90 |
|  | −z | 58 | 34 | 147.54 |
| Z | −x | 185 | 29 | 237.48 |
|  | y | 993 | 4 | 240.37 |
|  | −y | 394 | 15 | 264.95 |
| -X | −y | 101 | 50 | 222.24 |
| -Y | −x | 108 | 51 | 259.23 |
|  | z | 59 | 23 | 175.41 |
|  | −z | 59 | 23 | 175.35 |
| -Z | −x | 185 | 29 | 237.31 |
|  | y | 992 | 4 | 240.66 |
|  | −y | 167 | 35 | 262.49 |

The average speed of the surviving particles exhibited a dependence upon the direction of launch but only from the Y and –Y launch points. From the Y launch point there was a difference of 99.4 m/s in the average speed between particles which were initially traveling in the –x- and –z direction. For particles launched from the –Y point, particles initially traveling in both the +z or –z directions ended the simulation with a speed approximately 84 m/s slower than the particles launched in the –x direction.

Particles ejected at the poles also underwent a complex orbital evolution resulting in particle diffusion throughout their lifetime over their orbital path as illustrated in Figures 4 and 5. Particles launched with the same ejection speed but having different sizes diffused over time to follow the same general path but at slightly different speeds resulting in a "dotted-line effect". Similar effects were seen for particles launched from the –Z point (i.e. the "South Pole").

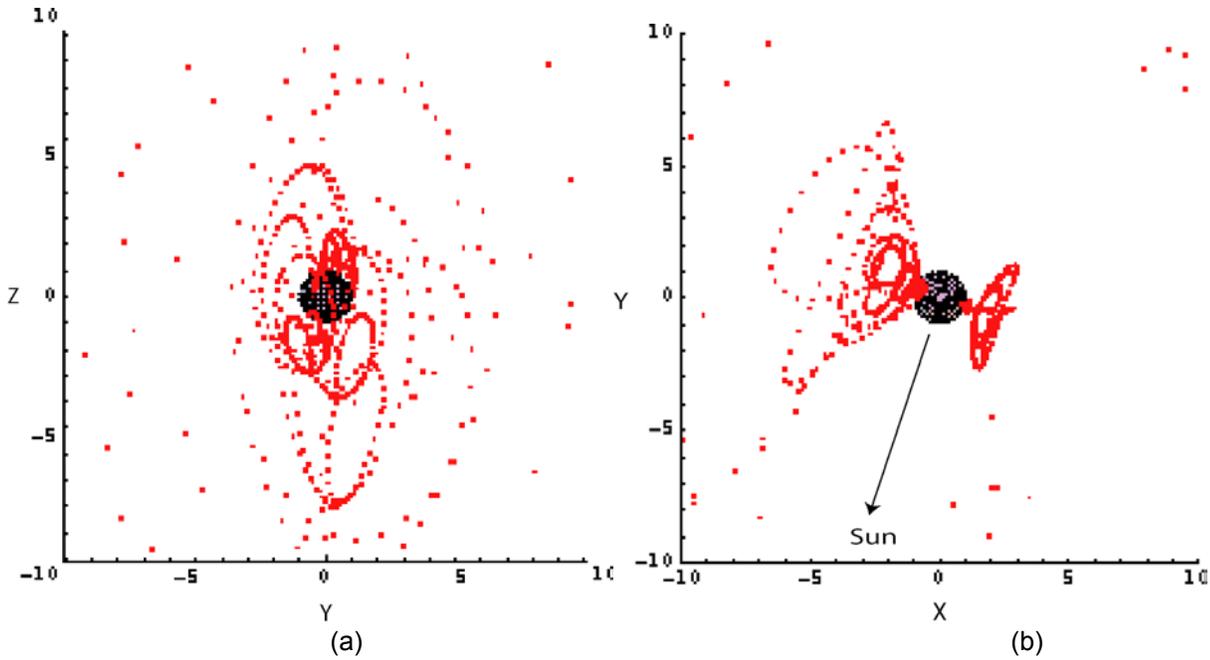

Fig. 4. Positions of the surviving particles (illustrated by light gray dots) launched from the +Z point and initially traveling in the +y direction after one Earth year. Axes are in Ceres radii. Figure 4 (a) illustrates a side view (along the +x-axis) while Figure 4 (b) illustrates a downward view (from above the orbital plane of Ceres, the +z-axis). Sun direction is indicated in Figure 4 (b).

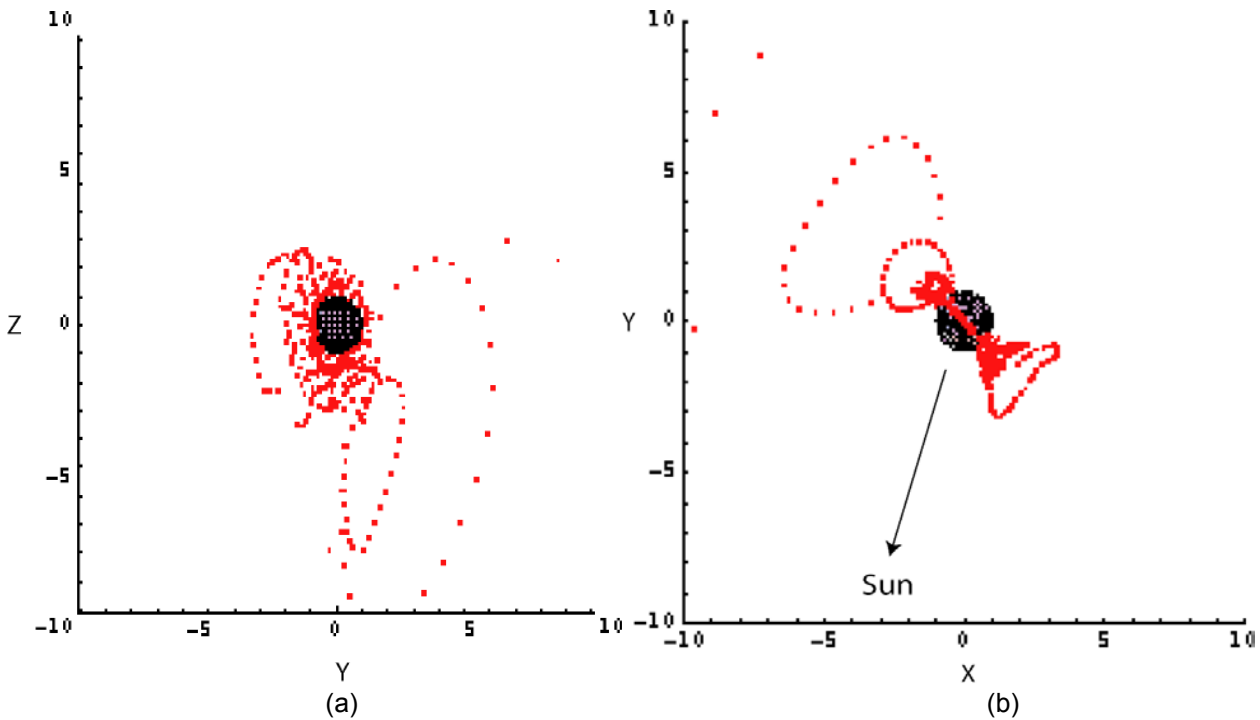

Fig. 5. Positions of the surviving particles (illustrated by light gray dots) launched from the +Z point and initially traveling in the -y direction after one Earth year. Axes are in Ceres radii. Figure 5 (a) illustrates a side view (along the +x-axis) while Figure 5 (b) illustrates a downward view (from above the orbital plane of Ceres, the +z-axis). Sun direction is indicated in Figure 5 (b).

## CONCLUSIONS

Previous studies investigating the fate of dust particles have primarily assumed they were already in orbit about an asteroid (Hamiliton and Burns, 1991, Hamiliton and Burns, 1992) or had been levitated from its surface (Lee, 1996). More recent studies (Nazzario and Hyde, 1997 and Nazzario, 2002) investigated a single impact point to determine that dust particles could survive at least one year in orbit around an asteroid such as Ceres. In contrast, this study investigated ejecta resulting from multiple impacts of small bodies onto an asteroid's (Ceres) surface. As shown above, the location of such an impact on an asteroid like Ceres will in large part determine the fate of the ejected particles. Particles ejected at the "poles" of the asteroid have the greatest probability of surviving for at least a one-year period even though the majority are quickly lost from the immediate vicinit of Ceres. The average particle speed around Ceres (for particles up to $50 \times 10^{-6}$ m) is low (< 300 m/s) but does exhibit variations depending on the position and direction of particle ejection. Even though the average speed is low, some dust particles are traveling faster than this and could pose a risk to spacecraft in the vicinity of an asteroid. The complicated orbital patterns of the dust ejected from the asteroid's "poles" leads to orbital stability zones having higher dust concentrations which might be observed by orbiting spacecraft. Also, the low average speed combined with the close proximity of the dust to Ceres will enable the remaining dust particles to enter orbits which may be stable for up to several years. This will be investigated in future work.